\documentclass[12pt]{iopart}

\usepackage{iopams}
\usepackage{dcolumn}
\usepackage{float}
\usepackage{graphicx}
\usepackage{color}

\definecolor{b}{rgb}{0.3,0.3,0.9}
\definecolor{g}{rgb}{0.3,0.9,0.3}
{\color{b} }

\def \sro{Sr$_2$RuO$_4$}
\def \tc{T$_{\rm c}$}

\def \musr {$\mu$SR}
\def \he3{$^3$He}

\begin{document}

\title{Chiral P-Wave Order in \sro}

\author{Catherine Kallin}

\address{Department of Physics and Astronomy, McMaster University, Hamilton ON L8S 4M1}

\ead{kallin@mcmaster.ca}

\begin{abstract}

Shortly after the discovery in 1994 of superconductivity in \sro, it was proposed on theoretical grounds that the superconducting 
state may have chiral p-wave symmetry analogous to the A phase of superfluid \he3.  Substantial experimental evidence has
since accumulated in favor of this pairing symmetry, including several interesting recent results related to broken time reversal symmetry
and vortices with half of the usual superconducting flux quantum.   Great interest surrounds the possibility of chiral p-wave
order in \sro, since this state may exhibit topological order analogous to that of a quantum Hall state, and can support such exotic physics 
as Majorana fermions and non-Abelian winding statistics, which have been proposed as one route to a quantum computer.
However, serious discrepancies remain in trying to connect the experimental results to theoretical predictions for chiral p-wave
order.  In this paper, I review a broad range of experiments on \sro\ that are sensitive to p-wave pairing, triplet superconductivity 
and time-reversal symmetry breaking and compare these experiments to each other and to theoretical predictions.  In this context, the evidence 
for triplet pairing is strong, although some puzzles remain.  The ``smoking gun'' experimental results for chiral p-wave, those which directly
look for evidence of broken time-reversal symmetry in the superconducting state of \sro, are most perplexing when the results are 
compared to each other and to theoretical predictions.  Consequently, 
the case for chiral p-wave in \sro\ remains unresolved, suggesting the need to consider either significant 
modifications to the standard chiral p-wave models or possible alternative pairing symmetries.  Recent ideas along these lines
are discussed.  
\end{abstract}

\maketitle

\section{Introduction} 

Explaining conventional superconductivity, as was done by Bardeen, Cooper, and Schrieffer (BCS) in 1957, was one of the major scientific accomplishments of the second half of the 20th century.\cite{bcs} The idea, conceived by Cooper, of an electron pairing instability for even a weak attractive interaction, quickly led to a revolutionary microscopic theory based on a new many-body wave function for the superconducting state.  In the particular model that BCS solved, the attraction resulted from interactions of electrons with lattice vibrations, and the resulting Cooper pair wave function, the superconducting order parameter or gap function, had s-wave symmetry.   By Fermi statistics, this spatially symmetric state of two electrons must combine with an antisymmetric spin singlet part. This form of BCS theory was extremely successful in describing the properties of all superconductors known at that time.  

Electrons interact primarily by the Coulomb interaction which has both a direct part, which is repulsive, and an exchange part, which can be attractive.  Higher angular momentum pairing wave functions, which vanish at short distance, minimize the direct Coulomb repulsion so that, in principle, p- or d-wave pairing, in which the electrons avoid direct overlap, could be stabilized by electron-electron interactions.  This case, where the gap function changes sign and averages to zero around the Fermi surface, is referred to as unconventional superconductivity.  Understanding unconventional superconductivity, both in general and in any specific material, is a major theoretical challenge.  Exactly how superconductivity results from the Coulomb interaction in a crystalline material can be quite complex and, in contrast to conventional BCS superconductivity, no controlled many-body perturbation method exists for treating this problem.\cite{unconv,norman}

Although there were hints of non-s-wave superconductivity in some of the heavy fermion superconductors, it was not until the discovery of high \tc\ superconductivity in the cuprates that such behavior was unambiguously identified.\cite{bednorz}  The high \tc\ cuprates were found to have d-wave symmetry, with lobes of the pair wave function along  $\pm x$ having the opposite sign as lobes along $\pm y$.\cite{hardy,dwave}  d-wave symmetry implies that the gap  vanishes at four nodal points on the two-dimensional Fermi surface relevant to these layered materials.  These gapless points give rise to power law behavior of low temperature thermal properties, in contrast to s-wave case where the low T behavior is exponential ($e^{-\Delta/T}$).  Interesting interference effects can also be observed because of the different phases of the different lobes of the pair wave function. Again for d-wave, the spins are paired in singlet states because of the even parity of the d pair wave function.

``What about p-wave?'', one might ask.  In fact, p-wave pairing was observed in the neutral superfluid, \he3, long before high \tc.\cite{lee-he3} The potential between two He atoms has a repulsive core, which works against s-wave pairing, and a weak, attractive van der Waals tail.  Two \he3 atoms, which like electrons are spin $1\over 2$ fermions, can take advantage of the attraction and avoid the repulsion by pairing in a p-wave state, and the correlations involved in doing this include ferromagnetic spin fluctuations. The spin states that go with p-wave pairing are triplet.   Since normal helium is an isotropic liquid, the three p-states, $p_x$, $p_y$, and $p_z$ are degenerate, as are the three triplet spin states.  In principle, the pairing wave function can be any linear combination of these, but not all linear combinations are equivalent.

Consider, for example, the $p_x$ state.  In $k$-space, the gap vanishes  for $k_x=0$ leaving a ring of gapless excitations on the Fermi surface.  Superconductivity is stabilized by the gap at the Fermi surface, and having nodes is costly in condensation energy.  The best situation is a fully gapped Fermi surface as happens for s-wave.  In fact, p-wave pairing can completely gap the Fermi surface.  The Balian and Werthamer (BW) state, in which the sum of the Cooper pair orbital and spin angular momenta is zero, has a gap which is uniform in magnitude around the  spherical Fermi surface of \he3.\cite{he3}  It describes the B phase, which is the lowest free energy state of superfluid \he3 in zero magnetic field, except for a tiny sliver of the pressure-temperature phase diagram at non-zero pressure on the boundary with the normal state.

Of more direct relevance to this paper is that tiny sliver, the A phase.  This phase picks out a direction, call this $z$.  In the A phase, the spin state is $S_z=0$ (i.e. the projection of  the triplet spin in the $z$ direction is zero) and the orbital part is of the complex form $p_x + i p_y$ or $p_x - i p_y$.  In other words, the A phase first breaks rotational symmetry by picking out the direction $z$, and it then breaks chiral symmetry by choosing a sense in which the phase of the gap rotates as one moves around the Fermi surface in a plane perpendicular to $z$.  Note that the magnitude of this chiral p-wave gap is not uniform around the Fermi sphere, since it vanishes at the north and south poles, the  directions $\pm z$, and is maximal at the equator.  However, a thin film of the A phase would be fully gapped. 

Superfluidity in \he3 is a fragile phenomenon, occurring at millikelvin temperatures, and so the question naturally arose whether a more robust form of triplet p-wave superconductivity, perhaps even chiral p-wave, could occur in a metal.  A possible candidate emerged with the discovery of superconductivity in \sro\  by the group of Maeno in 1994.\cite{maeno94}  \sro\ is a layered material and the Ru 4d-orbitals give rise to three bands crossing the Fermi energy.  As a result, the Fermi surface, which is almost perfectly cylindrical, has three sheets, one, an electron surface which is approximately circular in cross section, and two, an electron and a hole Fermi surface, which are more square with rounded corners. (See Fig.~\ref{arpes}.)  Since ferromagnetism is observed in closely related strontium ruthenate compounds, the idea that superconductivity in \sro\ might be triplet seemed appealing,\cite{rice} and, indeed, a number of experiments, which will be reviewed below, give evidence of triplet pairing and of the broken time-reversal symmetry (BTRS) associated with chiral p-wave superconductivity.

Putting aside for a moment the complications implied by a multi-sheet Fermi surface, one might ask which of the possible states of a triplet p-wave superconductor might be stable for \sro?  Several different linear combinations of the p-wave and triplet states fully gap a two-dimensional or cylindrical Fermi surface, including the two-dimensional analogs of the states describing the A and B phases of \he3.  Consequently, any of these would seem to be good candidates with a large condensation energy.  However, the chiral p-wave state differs from the BW-type states in that it breaks time-reversal symmetry.  It is also thought to have lower energy in the presence of spin-orbit coupling.\cite{ng}  Beside its relevance to superfluid \he3\ and \sro, chiral p-wave order is also of interest because it is an example of topological superconductivity. \cite{topsc1,topsc}

There is currently a great deal of interest in topological order which occurs in topological insulators, quantum Hall systems, superfluid \he3, as well as in chiral p-wave superconductors.\cite{topsc,toporder,topins}  All of these systems exhibit gaps in the bulk, but, because of the topologically ordered nature of the bulk, this gap must collapse to zero at a surface, giving rise to gapless surface modes.  For the case of topological insulators, the gapless modes are associated with spin currents.   For quantum Hall systems and chiral p-wave, the states in the gap propagate clockwise or counterclockwise around the edge, depending on the sign of the magnetic field for the quantum Hall case or of the chirality for chiral p-wave.   For chiral p-wave superconductors, these gapless modes are Majorana fermions, particles that are their own antiparticles and are like half of an ordinary fermion since a single electron occupies two Majorana modes.   Majorana zero modes are also expected to exist in the cores of superconducting chiral p-wave vortices.  In fact, such a superconductor may support vortices with half of the usual superconducting flux quantum, which obey non-Abelian winding statistics when moved around each other.\cite{nonabelian}  It has been noted that the extra stability associated with such topological states, in principle, could be used to minimize decoherence effects and hence find application as qubits for quantum computing.\cite{qc,qcomp,dassarma2006}

A large number of experiments have been performed on \sro, providing evidence both for triplet superconductivity and for BTRS and the combination of these strongly point toward chiral p-wave order.\cite{unconv}  Much of this paper is devoted to a critical analysis and comparison of the various results, focussing on experiments which most directly connect to evidence for triplet pairing or BTRS, with an emphasis on recent results.  To jump ahead to the conclusions of this analysis, although different experiments indeed show evidence of each of the key phenomena, the assumptions made in the analysis of different experiments are sometimes mutually incompatible, casting doubt on some of the conclusions.  Furthermore one key signature of chiral p-wave, spontaneous surface currents whose direction depends on the sign of the chirality, has been conspicuously absent in spite of repeated experimental efforts of increasingly high precision.  In the last two Sections of this paper, I suggest some avenues for future work and briefly discuss how a more thorough treatment of material-specific properties of \sro, such as spin-orbit coupling and the multi-sheet Fermi surface, could lead to alternative explanations of the different experimental results.

\section{A few key properties of \sro}

\begin{figure}
        \begin{center}
         \vspace*{0.0 cm}%
               \leavevmode
                \includegraphics[origin=c, angle=0, width=5cm, clip]{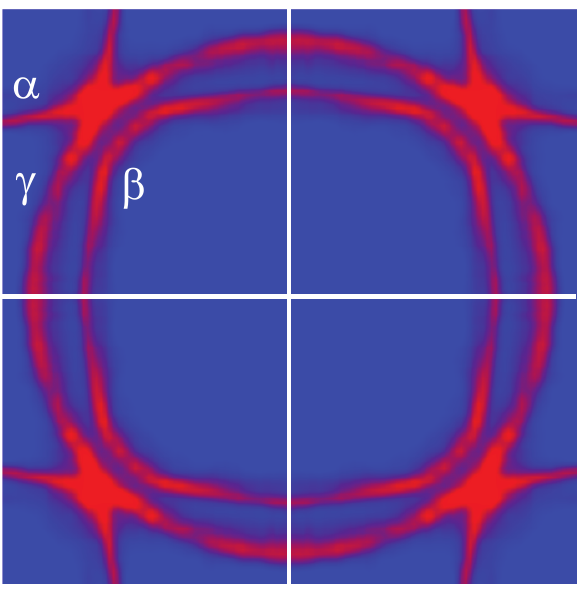}
         \vspace*{-0.5cm}%
        \end{center}
        \caption{The three sheets of the Fermi surface of \sro, labelled $\alpha$, $\beta$ and $\gamma$, are shown in the $(k_x,k_y)$ (or $ab$) plane as measured by angle resolved photoemission experiments.\cite{fs}  The $\Gamma$\ point (${\bf k}=0$) is at the center.  The dispersion in $k_z$ (not shown) is very small.  Figure provided by A. Damascelli.}
\label{arpes}
\end{figure}

\sro\ has the same tetragonal crystal structure as the cuprate superconductor, La$_2$CuO$_4$,\cite{mackenzie2003}  and, like the cuprates, is a highly anisotropic layered material, with good conduction in the RuO$_2$ layers (the tetragonal $ab$ plane, also taken to be the $xy$ plane here) and much weaker conduction along the $c$ (or $z$) axis, perpendicular to these planes.  However, other electronic properties of \sro\ are quite distinct from the cuprates.  While the cuprates, over much of their phase diagram, behave very differently from conventional metals described by Fermi liquid theory,  \sro\ behaves like a Fermi liquid below about 50K, although with substantial mass and susceptibility enhancements, signalling the existence of strong electronic correlations.\cite{mackenzie2003}  Furthermore, whereas the cuprates have a single band crossing the Fermi energy, \sro\ has three bands crossing the Fermi energy.  This is a key property which can significantly complicate the determination of the symmetry of the superconducting order.  The three sheets of the Fermi surface, labeled as $\alpha$, $\beta$ and $\gamma$, are shown in Fig.~\ref{arpes}.\cite{fs,dhva}  The Fermi sheets are shown for momenta in the $ab$ plane and there is very little dispersion along the $c$-axis, reflecting the weak conduction along the c-axis and the large anisotropy of this material.   The $\gamma$ band is mostly composed of Ru d$_{xy}$ orbitals, whereas the Ru d$_{xz}$ and d$_{yz }$ orbitals form one-dimensional bands which hybridize to give the $\alpha$ and $\beta$ bands.   The $\beta$ and $\gamma$ bands are electron-like and the $\alpha$ band is hole-like.

Another property which can complicate the identification of the symmetry of the superconducting order in \sro\ is the presence of strong spin-orbit coupling.   Detailed comparisons between the measured band structure and density functional theory calculations for this material reveal the effect of spin-orbit coupling on the band structure which is significant in the region of k-space where the 3 Fermi sheets are close to each other -- near the diagonals, $k_x=\pm k_y$.\cite{damascelli,mackenzie-so}  This analysis yields estimates for the atomic spin-orbit coupling of about 100 meV, or even larger,\cite{mackenzie-so} and also shows the effect of spin-orbit coupling to have significant $k_z$ dependence.\cite{damascelli}

\sro\ becomes superconducting below 1.5K, although the superconducting transition temperature, \tc, is very sensitive to non-magnetic impurities,\cite{mackenzie2003} as expected for unconventional superconductivity where scattering around the Fermi surface can average the superconducting order parameter to zero. 1.5K is the highest \tc\ observed.  Shortly after the discovery of superconductivity in \sro\ in 1994,\cite{maeno94} experiments found evidence both for triplet pairing and for broken time reversal symmetry (BTRS) as discussed below.   Triplet spin pairing implies that the spatial part of the pair wave function is odd. The simplest and, perhaps, most likely possibility is then p-wave pairing, since f-wave and higher angular momenta typically have higher energy, although this depends on the details of the microscopic pairing interaction.  Putting together the evidence for triplet pairing and BTRS, along with some energetic considerations, a symmetry analysis for the case of p-wave pairing and a cylindrical Fermi surface uniquely picks out a single pairing state - the chiral p-wave state.\cite{unconv,rice}  The electron-like $\gamma$ band is reasonably approximated by a cylindrical Fermi surface and small deviations from this simple approximation gives rise to gap anisotropy, but does not change the chiral p-wave state in a significant way.  There is indirect experimental evidence,\cite{deguchi} based on the variation of the low temperature specific heat with magnetic field direction, that points to the $\gamma$ band playing the dominant role in superconductivity, with weaker superconductivity on the $\alpha$ and $\beta$ bands.  Consequently, it is often assumed that one can largely understand the superconductivity within a one-band model, with substantially weaker superconductivity on the $\alpha$ and $\beta$ bands induced through the proximity effect.  

The simple picture of chiral p-wave is complicated by experimental evidence for nodes or deep minima in the superconducting gap seen in specific heat,\cite{deguchi,nishizaki} thermal conductivity,\cite{tanatar,tanatar2,izawa} nuclear spin relaxation,\cite{ishida2000} penetration depth,\cite{bonalde,baker} and ultrasonic attenuation measurements,\cite{lupien2001} which show power law behavior at low temperatures.  If these are accidental zeros or near zeros which do not correspond to sign changes in the gap function, then they are still compatible with chiral p-wave symmetry.  If there are nodes with a sign change, a crucial question is whether these are vertical, running along $k_z$, or whether they are horizontal, occurring at specific values of $k_z$.  Vertical nodes would be incompatible with chiral p-wave, whereas horizontal nodes could be compatible with some forms of chiral p-wave.  While it is possible to fit low temperature specific heat data with horizontal nodes, the detailed studies of the anisotropy of the specific heat in a magnetic field,\cite{deguchi} mentioned above, support line nodes or near nodes running along $k_z$.  One would like to see additional experiments addressing the issue of exactly where the low-lying excitations are in momentum space, as this connects directly to the issue of the symmetry of the superconducting order and, furthermore, could provide additional information on how active the different bands are in the superconducting state.  Although the work of Deguchi and coworkers\cite{deguchi}  provides important information which must be taken into account, both the issue of nodes or near nodes and the question of which bands are most important for superconductivity in \sro\ are still controversial.   I will return to the possibility of superconductivity on the quasi-one-dimensional $\alpha$ and $\beta$  bands later, but, first I will discuss the more commonly employed model of chiral p-wave order due to superconductivity primarily on the quasi-two-dimensional $\gamma$ band.

\section{Chiral p-wave superconductivity}

For a triplet superconductor, one needs to specify the pair wave function or order parameter for each of the three triplet states and this is typically done by expressing the order parameter in terms of a ``d-vector'', which rotates if the spins rotate. Each component of the d-vector represents the k-dependent gap amplitude for the spin state that has zero projection along the corresponding spatial direction.  In this notation, the BW state of the B phase of \he3\ is simply, ${\bf d}(p)=\Delta_0{\bf p}/p_F$, where $p_F$ is the Fermi momentum and $|{\bf d}|=\Delta_0$ is the magnitude of the gap around the Fermi sphere.  The analog of this state for a cylindrical Fermi surface has exactly the same form, but ${\bf p}$ is a two-component momentum perpendicular to the cylinder axis.  This state fully gaps the Fermi surface but does not break time-reversal symmetry. 

While there are many different p-wave states compatible with a cylindrical Fermi surface and tetragonal symmetry, only one of these, chiral p-wave, both breaks time reversal symmetry and is unitary.\cite{unconv,rice}  The non-unitary states, defined by nonzero ${\bf d}\times{\bf d}$, have different energy gaps for up and down spins and break time-reversal symmetry.  However, in zero magnetic field, the non-unitary states are expected to have a higher energy than the unitary states,\cite{zhit} and, consequently, the chiral p-wave state is singled out based on a combination of energetics and evidence of BTRS.

The d-vector describing the chiral p-wave order proposed for \sro\ is:
\begin{equation}
{\bf d}(p)=\Delta_0(p_x \pm ip_y){\bf \hat{z}}/p_F
\label{dvec} 
\end{equation}
where ${\bf \hat{z}} $ is the direction normal to the layers.  
The spin is zero along the direction of ${\bf d}$, so Eq.~(\ref{dvec}) describes Cooper pairs in an $S_z=0$ state with the orbital part winding by $\pm2\pi$ around the Fermi surface.   The $\pm$ corresponds to the two chiralities, which are degenerate and may coexist, although there is an energy cost associated with domain walls between the two chiralities.\cite{matsumoto}   As mentioned earlier the quasiparticle energy spectrum is fully gapped in the bulk, or everywhere for a system with periodic boundary conditions.  For a system with an edge, a chiral edge mode appears, which, splits off the bulk states below the gap and disperses across the gap, having exactly zero energy where the momentum, {\bf p}, along the edge is zero.\cite{read,stone}  The direction of this mode, whether it crosses from below the gap to above or vice versa, depends on the chirality of the gap function. The wave functions corresponding to these states in the gap are localized near the surface in a distance corresponding to the superconducting coherence length, and they carry a supercurrent along the edge. In addition, the bulk states, which are necessarily orthogonal to the surface states, also contribute to the surface current.\cite{stone}

The discussion so far, refers to states of a single layer.  For a 3D superconductor, consisting of a macroscopic stack of layers, the surface currents will generate a magnetic field at the surface which will be screened by diamagnetic screening currents of the bulk superconductor so that the net field inside the sample is zero.\cite{kwon}  Thus one expects a sheath of non-zero B-field at the surface which grows up over a coherence length and then falls to zero on the scale of the penetration depth.  The existence of surface currents is a robust consequence of the topology of the gap function, although the magnitude of these currents is not topologically protected.

It is also possible to have domains of positive and negative chirality in a single sample, in which case spontaneous supercurrents flow around each domain, giving rise to non-zero magnetic fields localized near the domain walls.  Domains are energetically costly because of the interfacial energy between domains.  Unlike in a ferromagnet, there is no compensating energetics driving domain formation.  However, provided there are inhomogeneities which can pin domain walls, domains nucleated at the superconducting transition can then be trapped in the sample at low temperature. 

The spontaneous edge currents and their associated magnetic fields are not expected to be a small effect in superconducting \sro.  Given the size of the gap, the Fermi velocity and the superfluid density, it is straightforward to estimate their magnitude and spatial distribution.  Expected values of the field are of the order of 10 Gauss over a region whose width is the coherence length plus penetration depth.\cite{matsumoto,kwon}  Such fields should be observable by scanning SQUID or Hall probes.

\section{Evidence for triplet pairing}

A key piece of information about the pairing symmetry is whether the electron spins in a Cooper pair form a singlet or a triplet.  A measurement of the spin susceptibility can distinguish between these two possibilities.  In a singlet superconductor, the spin susceptibility drops for any direction of applied field as singlet pairs are formed below \tc, whereas, for a triplet superconductor, the change in spin susceptibility from the metallic state depends on the particular triplet state and on the direction of the field.   The spin susceptibility contributes to the Knight shift which is measured by nuclear magnetic resonance (NMR) experiments. NMR experiments on \sro\, with the applied field in the $ab$-plane,  measure a Knight shift which is unchanged as the temperature is lowered through \tc.\cite{ishida1998}   The observed Knight shift is large and negative which implies that it is dominated by the electron spin susceptibility.  For a triplet state with $\langle S_z \rangle=0$, as in Eq.~(\ref{dvec}), the spins lie in the $xy$ or $ab$-plane and one expects no change in the spin susceptibility for applied fields in the $ab$-plane.  Therefore, the NMR data is taken as compelling evidence for triplet pairing.  Polarized neutron experiments also see no change in the spin susceptibility as the temperature is lowered below \tc.\cite{duffy}

More recently, nuclear quadrapole resonance (NQR) experiments measured the Knight shift in a field along $c$.\cite{nqr}  Again, to within the error bars, no drop in the Knight shift below \tc\ was observed .  This is not what is expected for the chiral p-wave state with the d-vector along $c$.   It has been proposed that the applied NQR fields, about 250G, may rotate the d-vector into the $ab$-plane, while remaining in the chiral p-wave state, as a similar effect is observed in He-3, where it is energetically favourable for the d-vector to be perpendicular to the applied field.  The pinning of the d-vector in the superconducting state is expected to be orders of magnitude weaker than the atomic spin-orbit coupling,\cite{ishida2008} so that fields of 250G may be sufficient to rotate the d-vector.
However, another possibility is that a field along the $c$-axis tilts the balance between different p-wave states, perhaps stabilizing the BW-like state, ${\bf d}\propto {\bf p}$, which has the d-vector lying in the $ab$-plane.  While a definitive interpretation of the NQR results is lacking, the NMR, NQR and polarized neutron data do not seem compatible with singlet pairing and seem to suggest a mostly triplet state that depends on the direction of the applied field.

In addition to looking for evidence for singlet versus triplet, one can also look for evidence of even versus odd parity pairing. This was done by connecting a conventional superconductor to \sro\ through two Josephson junctions on opposite $ac$\ faces of the \sro\ sample, and measuring the current as a function of magnetic field through the junctions.\cite{nelson2004}   The tunneling current between an s-wave and p-wave superconductor can be non-zero because of spin-orbit coupling.\cite{glb}  The current at zero field is expected to be a maximum (minimum) for even (odd) parity.  The observed signal is compatible with odd parity, implying triplet pairing, and is compatible with chiral p-wave order provided the entire sample is a single domain or, at least, only a few domains.\cite{nelson2004}  With many domains, one would expect the modulation with field to be greatly reduced, with maxima and minima at zero field occurring roughly equally as the measurement is repeated through successive cool downs. 

\section{Half Quantum Vortices}

New evidence in favor of chiral p-wave order, or more precisely, in favour of {\it two-component} triplet superconductivity, in \sro\ has come in the rather exotic form of evidence in favor of half quantum vortices (HQV).  Vortices in conventional superconductors are quantized in units of the superconductivity flux quantum, $\phi_0=hc/2e$, since this is the smallest flux compatible with the superconducting, or Cooper pair, wave function being single valued.  The phase of the wave function winds by $2\pi$ around the vortex, corresponding to the Bohm-Aharonov phase of a charge $2e$ Cooper pair encircling flux $\phi_0$.  However in a triplet superconductor, both the orbital and the spin part of the wave function can wind around the vortex.  For example, consider 
\begin{equation}
$$\psi=\Delta({\bf p})e^{i\theta/2}[-e^{i\theta/2}|\uparrow\uparrow\rangle + e^{-i\theta/2}|\downarrow\downarrow\rangle ] \ ,
\label{psi-vortex} 
\end{equation}
where $|\uparrow\uparrow\rangle$ is the state with both spins up along some direction, ${\hat n}$.   For $\theta=0$ and ${\hat n}={\hat x}$, the spin part of $\psi$\ is the $S_z=0$ state, and, for $\Delta({\bf p})=\Delta_0(p_x\pm ip_y)/p_F$, $\psi$\ is just the chiral p-wave order parameter.   If $\theta$ is the angle around a vortex, this wave function is single valued and has a winding of $2\pi$ for up-spins and no winding for down-spins.  In other words, it is a vortex only in one spin component.  Since only the orbital part couples to the magnetic flux and the orbital part only winds by $\pi$, this vortex corresponds to half of the usual superconducting flux quantum, or $\phi_0/2$.  

Chiral p-wave is not the only order which can support HQV since $\Delta({\bf p})$ in Eq.~\ref{psi-vortex} could be non-chiral -- for example, equal to $\Delta_0p_x/p_F$.  The essential feature for HQV is that the superconductor have two components, each of whose phase can wind.\cite{he3}  Chiral p-wave is one example of such a state, whereas the BW-type triplet state is not (although linear combinations of BW-type states may be).  So, some triplet superconductors (or superfluids) can support HQV.  However the energetics are such that HQV are not usually stable and, in fact, have never been observed in \he3.  Because HQV have both charge and spin currents, and only the charge current is screened, the energy of an isolated HQV grows logarithmically with system size.\cite{chung}  Therefore, one needs to look for HQV in samples that are not much larger than the penetration depth.  Even then, the energy of two HQV is typically expected to be higher than that of a single full vortex.  Since the spin part of the wave function in Eq.~\ref{psi-vortex} winds around the HQV, this implies the d-vector rotates around the HQV, which costs energy.  There is also an energy cost associated with suppressing the superconductivity at two half-quantum vortex cores relative to at one full vortex core.

Budakian and collaborators did ultra-high precision measurements of the magnetization of fabricated sub-micron sized \sro\ samples in an annular geometry, with an external field applied along $c$ (or $z$).\cite{budakian}  This is the classic flux quantization experiment, where the applied flux through the annulus plus the flux corresponding to the induced screening currents must sum to an integer number of flux quanta.  One can think of this as ``vortices", which sit in the center hole of the annulus, with their associated supercurrents flowing in the annular sample.  As expected, jumps in the magnetization, $\mu_z$, were observed as the field was varied, with the step height corresponding to integer flux quanta.  The surprising finding was that, when the experiment was repeated in the presence of an additional magnetic field in the $ab$-plane, these steps split into two half-steps.  The explanation provided in Ref.~\cite{budakian} connects to a theoretical prediction made concurrently with the experiments.  Vakaruk and Leggett pointed out that a HQV in a triplet superconductor necessarily has a spontaneous spin polarization.\cite{vakaryuk}  This is due to the fact that only one spin component is actively involved in the vortex, so that the superfluid velocity is spin dependent and, consequently, in equilibrium, there are different occupations or densities of the two spin components.  For an $S_z=0$ state, this polarization is in the $ab$-plane.  The interpretation of the experiment is then that this polarization couples linearly to the in-plane field, lowering the energy of the HQV with increasing in-plane field.  

This is a beautiful experiment with an exciting interpretation and, if the interpretation is correct, this is the first observation of HQV in the bulk of any superconductor.  Of course, one needs to be careful about other possible interpretations.  Budakian and collaborators have carefully addressed the more obvious alternatives, such as vortices entering the walls of the sample and exiting through the hole in the center, but, clearly, one would like to see this result independently confirmed.

These experiments provide further evidence for triplet pairing.  Together with other experiments, they strongly suggest chiral p-wave order, since the other unitary, two-component triplet states compatible with tetragonal symmetry do not break time-reversal symmetry.   If \sro\ is a topological chiral p-wave superconductor and if these are HQV formed under conditions where \sro\ remains in the chiral p-wave state, then there must be Majorana modes at the edge of the superconducting annulus.  One would then expect exotic features when two such HQV are moved around each other, since the HQV-Majorana fermion composites obey non-Abelian winding statistics.\cite{nonabelian}  It has been proposed that one might probe this physics through interference effects.\cite{interf}

\section{Broken time-reversal symmetry}

\begin{figure}
        \begin{center}
         \vspace*{0.0 cm}%
               \leavevmode
                \includegraphics[origin=c, angle=0, width=8cm, clip]{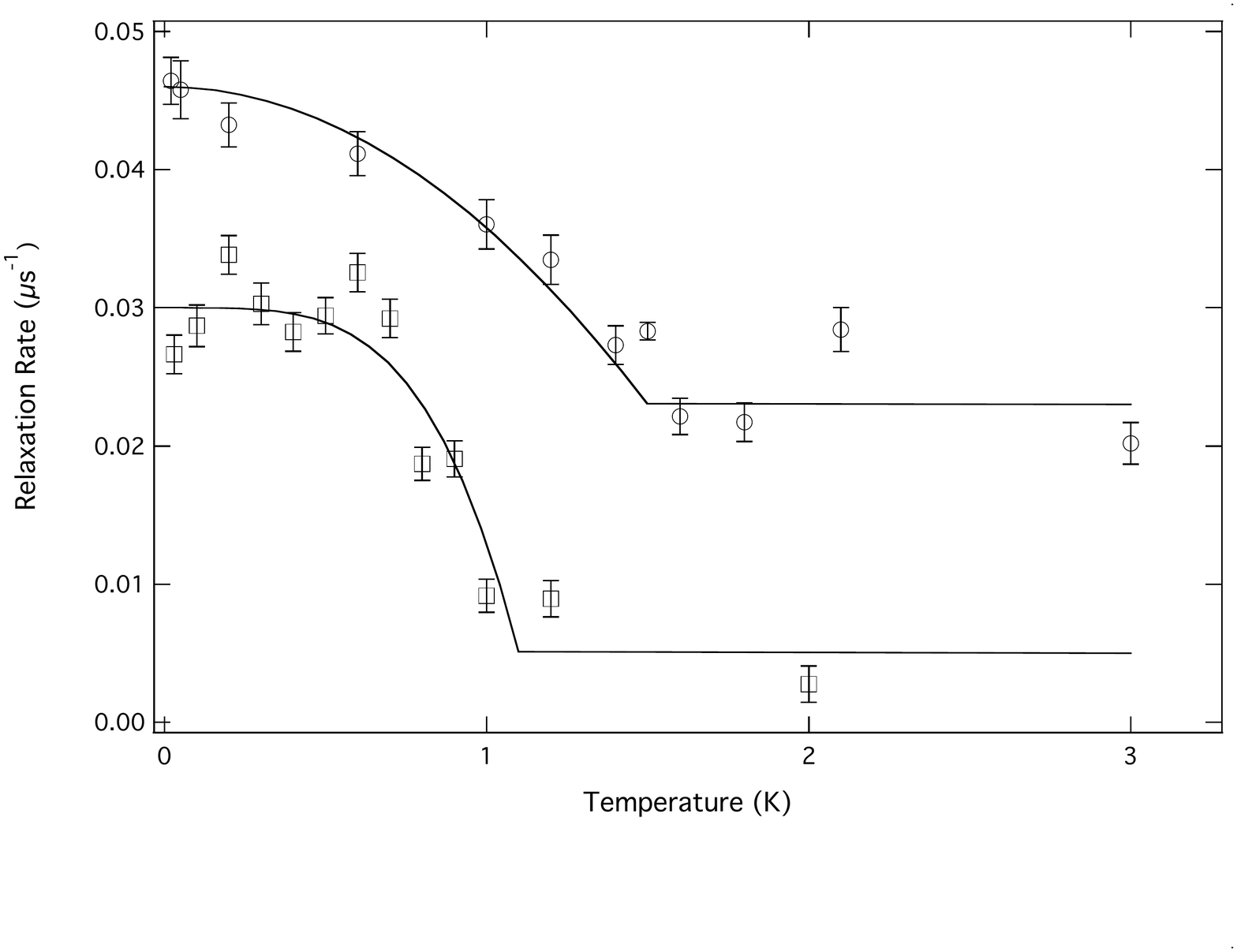}
         \vspace*{-1.2cm}%
        \end{center}
        \caption{The temperature dependence of the muon spin relaxation rate measured in zero magnetic field for two samples of \sro\ with \tc's of 1.45K (upper) and 1.1K (lower) is shown.  The lines are guides to the eye.  The extra relaxation below \tc\ indicates additional internal magnetic fields and, consequently, suggests the superconducting state has broken time-reversal symmetry.  Figure is from Ref.~\cite{luke2011}.}
\label{musrfig}
\end{figure}

Within a few years of the discovery of superconductivity in \sro, evidence that the superconducting state breaks time reversal symmetry was found in muon spin resonance (\musr) experiments, which probe local magnetic fields inside the sample.\cite{luke1998}  Fig.~\ref{musrfig} shows \musr\ data for two different \sro\ samples, with disorder added in one case to reduce \tc.\cite{luke2011}  The extra muon relaxation below \tc\ indicates additional internal fields and fits of the data are consistent with large but sparse magnetic fields that grow up below \tc.  While Meissner screening ensures that an ideal, single domain chiral p-wave superconductor has no spontaneous magnetic fields inside the bulk, magnetic fields are expected to exist near impurities, defects and domain walls.  No detailed chiral p-wave calculations exist which directly connect to the \musr\ data, but one can extract some general information from the data.  If it is assumed that the \musr\ signal is due completely to chiral p-wave domain walls, the fact that most muons see magnetic fields of a fraction of a Gauss or more suggests domains that, on average, are several microns or less in size.  This estimate of domain size is also consistent with the average field extracted from the muon data, assuming the field at domain walls is as predicted from the simple, single band chiral p-wave model.\cite{matsumoto}  More recent \musr\ data are consistent with this result,\cite{newmusr} and fits show that this data can also be well fit by assuming both domain walls and the muons themselves generate local fields.\cite{newmusr}  The fact that, in all cases, the extra muon relaxation
 associated with these internal fields turns on at \tc, even when \tc\ is reduced by the addition of impurities strongly suggests that the superconducting state has BTRS,\cite{luke2011} although one always has to be careful about alternative explanations.  For example, a change in lattice parameters as one cools through \tc\ (which has not been observed in \sro) could cause a change in the local fields due to nuclei.  

\begin{figure}
        \begin{center}
         \vspace*{0.0 cm}%
               \leavevmode
                \includegraphics[origin=c, angle=0, width=7cm, clip]{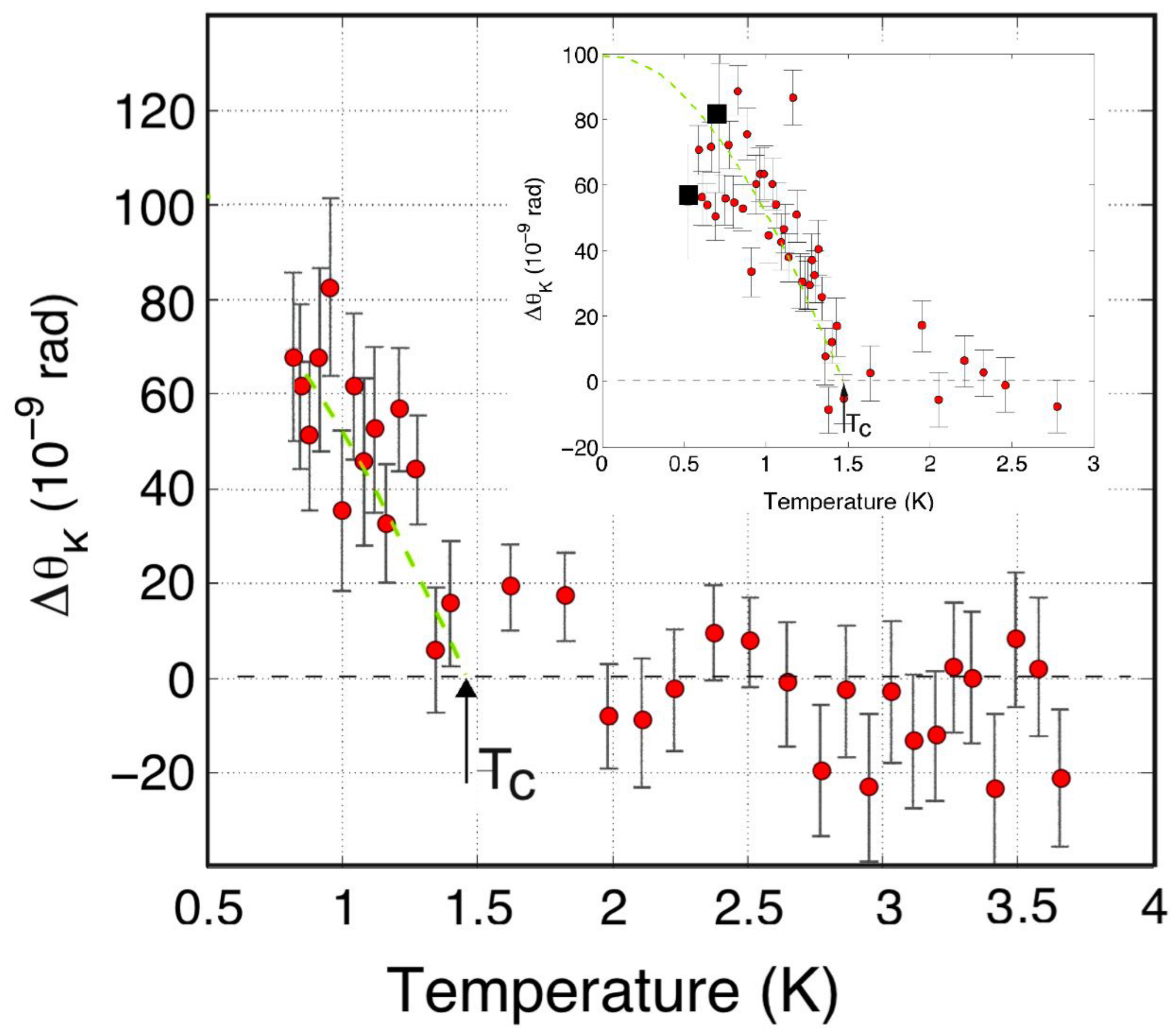}
         \vspace*{-0.5cm}%
        \end{center}
        \caption{The observed polar Kerr angle for \sro\  after zero field cooling. The inset shows data for a sample cooled in a field of 93 Gauss, followed by a zero field warm-up (circles).  The two solid squares are data taken just before the field was turned off.  Extrapolating the data to zero temperature, one estimates a Kerr angle of approximately 100 nrad. Figure taken from Ref.~\cite{kerr}.}
\label{kerrfig}
\end{figure}

Measurements of the polar Kerr effect, using a Sagnac interferometer, also provide direct evidence for BTRS in the superconducting state.  A nonzero Kerr angle, the relative angle of the polarization axis of the incident and reflected light, requires that the system preferentially absorb either right or left circularly polarized light.  Kapitulnik and collaborators observed a non-zero polar Kerr angle below \tc, as shown in Fig.~\ref{kerrfig}.\cite{xia2006,kerr}  The key features of the observed signal are that it is only non-zero below \tc, it increases roughly linearly with decreasing temperature close to \tc\ and it extrapolates to a zero temperature value of about 100 nrads at a probing frequency of 0.8eV. The sign and, to some extent, the magnitude of the observed Kerr angle are sensitive to cooling in a magnetic field.  In zero field cooling runs repeated over different regions of the sample,  both negative and positive Kerr angles were measured, sometimes with a reduced magnitude.  However, the maximum observed magnitude is consistent with that observed when the sample is cooled in a magnetic field, where the direction of the magnetic field selects the sign of the Kerr angle.  These features can be explained by the presence of domains that, for zero field cooling, are larger than but comparable to the beam size used to probe the sample.  Since the sign of the Kerr angle depends on the chirality its magnitude will be reduced if the beam averages over multiple domains.   The experimental beam size is about 25 microns, so domains that are roughly 50 microns across in the $ab$\ plane and that are large compared to the optical penetration depth (about 2000 \AA) along the c-axis, would be compatible with the observed Kerr effect.   On the other hand, the domains could be substantially larger than 50 microns in the $ab$\ plane if their extent along the $c$-axis is only comparable to the optical penetration depth.  A magnetic field can break the energy degeneracy between the two chiralities and favor larger domains of one chirality over the other chirality.\cite{bchir}  The Kerr effect experiments, when interpreted as evidence for chiral p-wave, imply that applied fields of less than 100G are sufficient to affect domain size and to create domains large compared to the beam size and depth to which it penetrates.

The polar Kerr experiment is strong, direct evidence of BTRS that turns on with the superconducting order and behaves qualitatively as expected for chiral p-wave domains. However, there is still a puzzle about the source of the effect.  Even though chiral p-wave breaks time-reversal symmetry, an ideal (one-band) chiral p-wave superconductor does not give a non-zero polar Kerr effect.\cite{read, royk}  This can be understood from the fact that the polar Kerr effect is, to a good approximation, proportional to the off-diagonal conductivity response, $\sigma_{xy}(\omega)$.  In a system with translational symmetry, $\sigma_{xy}$\ vanishes because the external field only couples to the center of mass momentum and, consequently, the conductivity is that of an ideal metal, {\it  i.e.}, a delta-function at $\omega=0$ in the real part of the diagonal conductivity.  Therefore, one needs to explicitly invoke a mechanism which breaks translation symmetry to obtain a non-zero $\sigma_{xy}$ or Kerr angle in a chiral p-wave superconductor and the magnitude of the Kerr angle will depend on the magnitude of the translational symmetry breaking.  

Until very recently, the most promising explanation that reconciled the Kerr effect experiments with chiral p-wave order involved impurity scattering.\cite{goryo}   This effect is expected to be rather small since \sro\ is necessarily in the clean limit in order to achieve a \tc\ of 1.5K and, consequently, the electron scattering rate due to impurities cannot be too large.   Furthermore, the usual dominant Born contribution to impurity scattering, which is proportional to $n_i \langle U^2\rangle$, where $n_i$ is the the density of impurities and $U$ is the impurity potential,  gives a tiny contribution to $\sigma_{xy}$ because it also requires particle-hole asymmetry.\cite{lutchyn}  The dominant impurity contribution comes from ``skew-scattering'' terms proportional to $n_i \langle U^3\rangle$.\cite{goryo, lutchyn}   At the experimental probing frequency, the Kerr angle depends rather sensitively on parameters which are not well known, so it is still an open question as to how well this model describes the experiment.\cite{goryounpub}  Experiments on purposely disordered samples would help determine whether disorder is playing the key role.  Recently, it was shown that a multi-band chiral p-wave superconductor can give rise to a nonzero Kerr effect, even in the absence of disorder.\cite{taylor}  For \sro, this effect requires substantial superconductivity on the $\alpha$\ and $\beta$\ bands and is discussed below, in Sec.~8, in connection with multi-band models of superconductivity.

Another experiment which can give information about TRSB is Josephson tunnelling between a conventional superconductor and \sro.  A configuration which gave information about the parity of the superconducting order was discussed above under triplet pairing.\cite{nelson2004}  In a corner junction geometry, the two junctions connect an $ac$\ face to a $bc$\ face, thereby probing the phase change in the \sro\ order parameter associated with a change in angle of $\pi/2$.  For chiral p-wave this would be probing the $\pm i$, or, in other words, directly probing the TRSB of the order parameter. The magnetic field dependence expected for a phase change of $\pi/2$ was seen in one such corner junction measurement, whereas more complicated field dependence was seen in other corner junction measurements.\cite{nelson2004}  On one hand, it is difficult to imagine how the expected signature for $\pi/2$ would be generated except through an order parameter with a phase structure like chiral p-wave.  On the other hand, the fact that this signature was not reproduced, leaves the corner junction signature somewhat open.  Certainly domains could complicate this experiment and possibly explain the results.  However, the same experiment in another geometry yielded evidence for odd-parity and appears not to have been complicated by domains.\cite{nelson2004}

One can also put the two junctions on the same $ac$ face of the crystal.  In the absence of domains, for any pairing symmetry, one would expect to see the classic Franhoefer pattern, peaked at zero field, corresponding to no phase change of the superconductor between the two junctions.   Instead, quite a complicated pattern with applied field was observed, but one which can be qualitatively replicated within a chiral p-wave model, if one assumes small domains, of the order of 1 micron, which intersect the crystal surface at oblique or acute angles (not 90$^\circ$).\cite{kidwingara2006,bouhon}  These domains must also be dynamic, as features in the pattern changed on the timescale of several seconds.  In principle, strong surface disorder could partially pin a dense array of domains near the surface.

\section{Where are the surface currents?}

\begin{figure}
        \begin{center}
         \vspace*{-2.6 cm}%
               \leavevmode
                \includegraphics[origin=c, angle=0, width=9cm, clip]{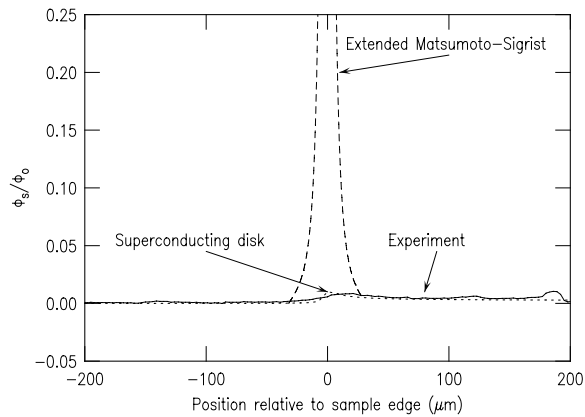}
         \vspace*{-3.4cm}%
        \end{center}
        \caption{The solid line shows the observed magnetic flux from a SQUID scan across the edge of an ab face of a \sro\ crystal.  The dotted line is the prediction for an s-wave superconducting disk in a uniform residual field of 3 nT and the dashed line (whose peak value is 1) is the prediction for a single domain chiral p-wave superconductor, following the theory Matsumoto and Sigrist,\cite{matsumoto} but modifed for a finite sample. From Kirtley et al., Ref.~\cite{kirtley}.}
\label{edgecurrents}
\end{figure}

A striking consequence of chiral p-wave order is the existence of spontaneous supercurrents at sample edges and at domain walls.\cite{matsumoto,stone,kwon}  Although the \musr\ data is interpreted as evidence for such currents, direct searches for these supercurrents at sample edges and surfaces, as well as in micron size samples, have yielded null results.   Scanning SQUID and Hall bar probes\cite{hall,kirtley,hicks} and cantilever magnetometry\cite{budakian} have been employed in the search for a signal due to spontaneous supercurrents.  Fig.~\ref{edgecurrents} shows the result of one such experiment, comparing the measured magnetic flux through a SQUID pickup loop as it is scanned across the edge of a \sro\ crystal to the theoretical prediction for a simple chiral p-wave model appropriate for superconductivity on the $\gamma$ band.  These experiments, as well as more recent scanning SQUID experiments,\cite{hicks} have placed a bound on the maximum value of the magnetization at the edge or at a domain wall intersecting the $ab$ surface that is three orders of magnitude smaller than  predicted, assuming domains larger than a few microns in size.  Alternatively, if the magnetization is of the predicted size, then these experiments would imply that the domains are 300 \AA\  or less in size for a random arrangement of domains.  The magnetometry measurements were done on micron sized samples and also conclude that the spontaneous currents need to be reduced by roughly three orders of magnitude if the sample were a single domain of chiral p-wave.\cite{budakian}  Given the small sample size, this null result would also require surprisingly small domain sizes, as a few random domains would not so precisely cancel the expected signal.  On the other hand, the Kerr effect\cite{xia2006,kerr} and odd-parity tunneling\cite{nelson2004} experiments require domains which are substantially larger than a few microns in size in order to be compatible with chiral p-wave order.  Taken together, these experiments suggests that if \sro\  is a chiral p-wave superconductor, the spontaneous supercurrents are dramatically smaller than expected for the simple chiral p-wave model.  

Disorder, band anisotropy, scattering at the surface, or other surface effects such as nucleating a competing order parameter near the surface, can all reduce the magnitude of the spontaneous currents at the edge and, in some cases, at domain walls. These effects have been studied with Ginzburg-Landau theory, and shown to typically give only a modest change in the magnitude of the spontaneous currents for a one-band model and Ginzburg-Landau parameters which are physically plausible.\cite{ashby}  Nucleating a different order at the surface might explain the null scanning results if this surface order persisted to the penetration depth, but this would then leave other experiments unexplained, such as the Kerr effect in the case of a (001) surface or tunnelling in the case of (100) surfaces.  In the special case of chiral p-wave order and complete retroflection at the surface, the edge currents can be fully suppressed.\cite{sauls-surface}  The more typical case of rough surfaces and diffuse scattering may cause more significant changes in the currents in the case of multiband superconductivity and this is discussed in the next section.  
                                                                                                                                            
\section{Can experiments be reconciled with chiral p-wave order?}

The evidence that \sro\ is an unconventional superconductor with triplet pairing is fairly strong.   However, the case for chiral p-wave order remains murky, in large part because the evidence for BTRS either does not directly connect to calculations for chiral p-wave or requires conflicting and/or special assumptions about chiral domains.\cite{kb} 

One of the more striking discrepancies between experiment and the assumption of chiral p-wave order is the multiple null results for observable effects due to spontaneous supercurrents.  In principle, the null results could be explained by sufficiently small domain size since they are either probed with a finite size pickup loop, as in the scanning Hall bar and SQUID measurements,\cite{kirtley,hicks} or averaged over the entire sample, as in the cantilever magnetometry measurements.\cite{budakian}  However, the experiments now put quite stringent bounds on domain size as discussed above.  Not only are such small domains incompatible with the interpretation of the Kerr and tunnelling experiments, but they would also be surprising since there is no identified energetic driving force for domain formation, as there is in a ferromagnet, and the samples are in the clean limit.  An ideal, defect-free crystal cooled sufficiently slowly through \tc\ is expected to be a single domain.  Hicks {\it et al.} discuss the possibility of spatially periodic domains which can be substantially larger and still escape detection as the resultant fields at the surface would be noticeably smaller than for spatially random domains of similar size.\cite{hicks}  This possibility may not be ruled out by existing experiments, but, in the absence of some energetic driving force that favors domain formation, one would not expect periodic domains.

In a single band model of chiral p-wave, it is difficult to escape having substantial edge currents at low temperatures.  A significant fraction of the edge current is provided by the edge modes whose key features are topologically protected, {\it i.e.},  they are chiral with a linear dispersion at low energy.  While the current can be reduced by disorder or band anisotropy, these effects are generally relatively small since the chiral nature of the edge modes implies that they cannot be localized by disorder.  The special case of perfect retroflective surface scattering does have a dramatic effect, driving the edge mode dispersion to zero. \cite{sauls-surface}  This special case is not likely to apply to \sro\ samples and, in any case, would not explain the lack of any observable currents due to domain walls intersection the (100) surface.  In the absence of disorder and in a simple geometry (e.g. a disk), the magnitude of the spontaneous edge current (which is screened in a charged superconductor) corresponds to the macroscopic angular momentum, $L=N\hbar/2$, where $N$ is the total number of conduction electrons.  This macroscopic angular momentum has been extensively discussed in the literature as the ``angular momentum paradox".\cite{leggett}  In the strong coupling limit, all conduction electrons condense into non-overlapping pairs, each with angular momentum $\hbar$, and $L=N\hbar/2$ follows immediately.  It may seem counterintuitive that for an arbitrarily weak pairing interaction, where only a tiny fraction of electrons are paired, the same macroscopic angular momentum characterizes the condensed state.  In fact, predictions for the magnitude of  $L$ in an ideal chiral p-wave state have ranged over six orders of magnitude.\cite{leggett}  Most of these predictions preceded our understanding of the topological nature of the chiral p-wave state, and the macroscopic angular momentum for the ideal case appears to no longer be in dispute, at least within the BCS formalism.\cite{stone,read}  Leggett has suggested that if one moves beyond the BCS or Bogliubov-de Gennes formalism used to describe the chiral p-wave state, a reduction by orders of magnitude may be possible.\cite{leggett}  While such a suggestion is intriguing, given the lack of experimental evidence for edge currents, a competing theoretical framework would necessarily have to explicitly treat an edge or boundary in order to have a well-defined $L$, and this is a formidable task when one moves beyond the simplicity of the BCS paired state.   There is no uncertainty that the A phase of \he3 is chiral p-wave and, consequently, that it is expected to exhibit this macroscopic angular momentum when confined to a thin disk.  However, it is a difficult experimental challenge to detect this angular momentum or mass flow in a neutral system, and direct confirmation of this property in \he3 is also still an outstanding problem. 

Unlike \he3, \sro\ has multiple bands near the Fermi energy, which can significantly change the predictions made for a single band model.  The one-band model is an approximation for the case where superconductivity is predominantly on the $\gamma$ band, with weak, induced superconductivity on the $\alpha$\ and $\beta$\ bands.  If all three bands are gapped, many of the properties discussed above, including the existence of substantial edge currents, are not expected to be significantly changed by including the effect of all three bands. On the other hand, if nodes or near nodes are present, as the experiments imply\cite{tanatar}-\cite{lupien2001}, this could noticeably change the prediction for edge currents by mixing the chiral edge modes with other low-lying states in the presence of disorder, although this effect has not been studied in any detail.  

Models with superconductivity primarily on the quasi-one-dimensional $\alpha$\ and $\beta$\ bands have also been studied,\cite{1dbands} and this possibility has recently been given prominence by a renormalization group approach which predicts chiral p-wave order on the quasi-1d bands in the weak coupling limit.\cite{raghu}  This changes things considerably, since the $\alpha$\ band is electron-like while the $\beta$\ band is hole-like.  The topological invariant which characterizes chiral p-wave order depends not only on the chirality, but also on the sign of the charge carriers.  Therefore the net topological invariant for these two bands is zero, which means the state is topologically trivial.\cite{raghu}  Although there are still Andreev bound states at the edge, they are non-chiral and not topologically protected.  Since the edge modes are quasi-one-dimensional, they will be localized by disorder, causing a noticeable reduction in the edge currents.  In general, the bulk states also contribute to the edge currents, but this contribution is expected to be noticeably smaller than it is for superconductivity on the $\gamma$ band since it depends on interorbital mixing and it may also be further reduced if the edge modes are localized by disorder.  

 Even if the pairing interaction only acts on the  $\alpha$\ and $\beta$\ bands, one expects superconductivity to be induced on the $\gamma$\ band, although this is expected to be weak because the $\gamma$\ band does not mix with the other two bands in the absence of spin-orbit coupling.  In this case, one might need to go to quite low temperatures to see superconductivity developed on all three bands.  At sufficiently low temperatures, however, one would still expect large edge currents, since the net topological invariant for all three bands is nonzero, giving rise to one chiral edge mode.  Again, the caveat about mixing of these edge modes with low-lying bulk excitations would apply, since the quasi-1d theory predicts deep minima or near nodes in the gap along the (1,$\pm$1) directions.\cite{raghu}
 
The quasi-1d model for superconducting \sro\ is an interesting proposal which could explain some of the experimental puzzles.  This model very naturally gives rise to deep minima or near nodes in the superconducting gap and, so, would give rise to power-law behavior at the low temperatures probed by experiments.   In addition, this model is likely to yield substantially reduced spontaneous currents at the temperatures studied by existing experiments, although this is yet to be verified by detailed calculations.  There is no reason to expect the Kerr effect to be similarly reduced, since the Kerr effect more directly probes the chirality of quasiparticle excitations and not the local fields induced by supercurrents.  In fact, recent work has shown that this model can give rise to a substantial Kerr effect even in the absence of any disorder.\cite{taylor,annett2}  The size of this intrinsic Kerr effect is proportional to the square of the maximum gap on the $\alpha$ and $\beta$ bands, while the $\gamma$ band only pays an indirect role with little effect on the Kerr angle.\cite{taylor}  Experiments on purposely disordered samples should be able to determine whether this intrinsic effect or the previously identified disorder effect dominates.  If the intrinsic effect is found to dominate in \sro, this would be compelling evidence for substantial chiral superconductivity on the quasi-1d bands.

Both classes of models discussed, the $\gamma$ band model and the quasi-1d model, exhibit chiral p-wave order described by the order parameter of Eq.~\ref{dvec}, although more accurately the gap function, $\Delta({\bf p})$, would have a band index and an anisotropic magnitude around the Fermi surface.  Furthermore, for strong spin-orbit coupling,  the direction of the d-vector should be thought of as specifying a pseudospin triplet state.   Other possible triplet order parameters include the BW-like states, of which there are actually four, ${\bf d}\propto (p_x {\hat x}\pm p_y{\hat y}), (k_y{\hat x}\pm k_x{\hat y})$.  These states are all fully gapped, but expected to lie somewhat higher in energy than the chiral state in the presence of spin-orbit coupling.\cite{zhit}  None of these states have BTRS, although one can form a linear combinations of these which do exhibit BTRS and which correspond to chiral p-wave with the d-vector lying in the $xy$ plane.  Since these four states are not expected to be degenerate in the presence of general spin-orbit interactions, it seems unlikely that this linear combination would be a ground state in zero magnetic field.   More generally, depending on the microscopic pairing mechanism, the superconducting order parameter for \sro\ may not be simply characterized by s, p or d-wave symmetry, and may mix singlet and triplet states, because of the combined effects of multiple bands and spin-orbit coupling.  For example, Puetter and Kee\cite{kee} recently included multi-band effects and spin-orbit coupling in a model where the pairing mechanism is the Hunds coupling between different $t_{2g}$ orbitals on the same ion.  Although time reversal symmetry is not broken in this state, there is a non-trivial phase relation between different components of the gap function which could lead to interesting interference effects.   Consequently, this model does not describe the \musr\ and polar Kerr measurements, but could explain the Josephson junction experiments which are also taken as evidence of BTRS.  More generally, this model shows some of the complexity that can arise even within an onsite pairing model and, like the quasi-one-dimensional model, it also can give rise to near nodes.  Puetter and Kee argue that the addition of nearest neighbor pairing interactions to this model could bring in BTRS and signatures of that have been identified with chiral p-wave order.

 \section{Conclusions and Future Directions}
 
So what can we conclude about whether or not \sro\ is a chiral p-wave superconductor with an order parameter similar to that of Eq.~(\ref{dvec})?   If one takes the evidence for triplet pairing at face value, the key experiments which favor a positive conclusion are the spontaneous appearance of random magnetic fields below \tc, observed in \musr, the observation of a polar Kerr effect below \tc, and a variety of interference patterns detected by Josephson tunnelling. 

Firstly, the \musr\ results are attributed to spontaneous currents generated at domain walls and defects.  This interpretation is undercut by the distinct absence of spontaneous surface currents which would be expected to occur for identical reasons.  Secondly, the observation of a polar Kerr effect would be more compelling as evidence for chiral p-wave if more detailed connections between theory and experiment could be made, both in the magnitude of the effect and in the role of disorder.  The recent prediction of an intrinsic Kerr effect due to superconductivity on the quasi-1d bands opens up an alternative explanation for the existing experiments and gives additional impetus for further experimental studies.  As mentioned previously, detailed comparisons between theory and experiment on purposely disordered samples might shed light on the important question of which bands are primarily responsible for the superconductivity.  Thirdly, the interference effects observed by Josephson coupling require a fairly specific domain pattern to be understood within the chiral p-wave model.  They suggest that some kind of complicated phase relationships occur between different points on the crystal surface, but the link between this and chiral p-wave is not straightforward.  

In my opinion, the mechanism or mechanisms which give rise to the  \musr, polar Kerr effect, and Josephson interference effects observed in \sro, in other words all the experiments directly connected to time-reversal symmetry breaking, are not sufficiently understood.  Information on the local magnetic fields as a function of the depth into the sample, such as might come from slow muons or $\beta$-NMR, for example, could, in principle, help resolve the puzzle of reconciling the null scanning probe results with the positive \musr\ results.  Detailed calculations of the local fields associated with various chiral p-wave domain patterns could make closer contact to the \musr\ results, much as is done in extracting information from \musr\ in the mixed state of \sro\ and other superconductors.  Finally, direct observation of domain walls would be most interesting.  The experiments which probe TRSB need to make specific assumptions about the existence of domain walls which are not mutually compatible with each other and with other experiments as they assume domain sizes which differ by up to 4 or 5 orders of magnitude in size.  Different samples and different cooling rates and residual fields can affect domain size and some attempt has been made to study possible domain alignment with magnetic field cooling in \sro\ Josephson junctions.\cite{domains}  Additional studies which attempt to directly control and observe possible domain walls in \sro\ would be very useful.

Futhermore, there is an important feature that is clearly observed in superconducting \sro\ which has implications for the chiral p-wave picture.  That is the obvious presence of line nodes or near nodes in the superconducting gap.  Although horizontal nodes would be compatible with chiral p-wave, it is not clear why they would arise in this highly layered material, and, as noted above, vertical nodes with a sign change are not compatible with chiral p-wave.  Near nodes, or deep gap minima, along the $k_z$ direction, do arise within some multiband models,\cite{raghu,kee} which suggests that further experiments which address the location in momentum space of the lowest lying excitations could provide key information on the order parameter symmetry and on which bands are participating.  Angle resolved photoemission is the obvious probe for this information and the technique has been very successful in providing momentum dependent information on the superconducting gap and low-lying excitations in the curates and pnictides, but it does not presently have the required resolution to address the same issue in \sro, where weak coupling BCS theory would predict a maximum gap of only 0.23eV.

A number of probes can offer less direct evidence of which bands are participating in the superconductivity.  The already mentioned specific heat measurements in a magnetic field are interpreted as evidence that the superconductivity is primarily on the $\gamma$ band, not the quasi-1d bands.\cite{deguchi}  In addition, inelastic neutron scattering studies found no change in the magnetic response below \tc.\cite{braden}  Since nesting of the $\alpha$\ and $\beta$\ bands is most likely responsible for the strong antiferromagnetic fluctuations observed, this study might suggest that no significant gap opens on these quasi-1d bands.  However, a weak coupling gap would be near the limit of the energy resolution for these experiments and the maximal gap is not predicted to occur at the wave vectors where the antiferromagnetic fluctuations are strongest.  Higher resolution at low energies would be needed to fully address this issue.  Tunneling has the advantage that it has the resolution to detect very small energy gaps, but as already noted, tunnelling data on SRO is often not simple to interpret and may involve poorly understood matrix elements.  Interestingly,  c-axis tunneling on \sro\ shows a full gap, with no low-lying density of states, and with a gap magnitude of 0.28meV, fairly close to the weak coupling BCS value.\cite{suderow}  Since the $\gamma$\ band has much weaker c-axis dispersion than the quasi-1d bands, one would expect the c-axis tunneling to be dominated by the quasi-1d bands.  If this is the case, this experiment lends support to the quasi-1d model.  In-plane ($ab$) tunnelling data has been interpreted as evidence for chiral p-wave edge modes.\cite{abtunnel}  However, the features which are expected to signal the gap magnitude are observed at 0.93eV, about 4 times larger than the BCS value, and this energy scale varies little with temperature below \tc, making it difficult to unambiguously interpret the results.  Nevertheless, all these different experiments are examples of the type of experiments which can, in principle, yield new insights into the nature of the superconducting order and, in particular, into which bands are participating.  

On the theoretical side, recent work has provided new insights into multiband pairing and spin-orbit effects.\cite{raghu,kee,taylor,annett2}  For example, the quasi-one-dimensional model might resolve the puzzles associated with BTRS and with nodes within a model of chiral p-wave pairing.\cite{raghu}  However, in that case, it would still appear that the \musr\ would need an alternative explanation.  It also seems that multiband and spin-orbit effects are required to understand the combined NMR and NQR data, discussed earlier, which shows no drop in the Knight shift at low temperatures for any field orientation. 

In summary, many puzzles remain regarding the symmetry of the superconducting order parameter in \sro.  Meanwhile, experiments on \sro\ continue to provide us with surprises, such a the recent magnetometry measurements taken as evidence of HQV.\cite{budakian}  More experiments of the types described here and new theoretical work, particularly on multiband models that include spin-orbit coupling, are needed to directly address the puzzles highlighted in this paper and to unambiguously determine whether the order is chiral p-wave.   However, in the meantime, one can still take the approach of forging ahead under the assumption of chiral p-wave order, possibly exploiting the geometry used in the magnetometry measurements to see half flux jumps, and looking directly for some of the more exotic physics associated with Majorana fermions and non-Abelian winding statistics.

\section{Acknowledgements.}

I have benefitted from comments and discussions with many colleagues, particularly with John Berlinsky,  Raffi Budakian, Clifford Hicks, John Kirtley, Graeme Luke, Yoshi Maeno, Kathryn Moler, and Manfred Sigrist.   This work was supported in part by the Natural Science and Engineering Research Council of Canada and by the Canadian Institute for Advanced Research.

\eject

\end{document}